\documentclass[12pt]{article}
\usepackage{subfigure}
\usepackage{epsfig} 
\begin{document}
\thispagestyle{empty}
\rightline{EFI 97-13}

\begin{center} \large {\bf High Energy Hadronic Total Cross-Sections}

\medskip\normalsize 
James A. Feigenbaum, Peter G.O. Freund and Mircea Pigli\\

             {\em Enrico Fermi Institute and Department of Physics\\
             The University of Chicago, Chicago, IL 60637, USA}

 \end{center}
 \bigskip
 \bigskip
\bigskip \bigskip \bigskip

\noindent{\bf Abstract}: High energy hadronic total cross-section data 
are found to agree with the predictions of a QCD-string picture.
\setcounter{page}{0}

\newpage

\bigskip

{\bf 1. Introduction}
\bigskip

High energy hadron-hadron scattering amplitudes at fixed momentum transfer
have two components: a diffractive component (Pomeron exchange) and a mesonic 
Regge exchange component. Since the latter decreases with energy, Froissart's
unitarity bound becomes a constraint on the diffractive 
component:
at high energies hadronic total cross-sections cannot exceed 
$\frac{\pi}{m_{\pi}^2} \ln^{2}s$. Although fits to total cross-section data
incorporating $\ln s$ and $\ln^2 s$ terms 
have been made, it has been noted 
that the unitarity bound does not rule out a term of the form 
$X (\frac{s}{m^2})^\epsilon$
with small but positive $\epsilon$. Indeed, for say $\epsilon = 0.08,~
m\approx 1$ GeV and 
$X \leq 25$ mb, the unitarity bound violation would not set in \cite{PDB} until energies 
of $10^{24}$ GeV are reached. At such super-Planckian energies the theoretical 
underpinnings of QCD become meaningless and experiments become impossible.
With this in mind, Donnachie and Landshoff \cite{DL} have  successfully fit
high energy hadronic total cross-sections to expressions of the type
$$
\sigma_{ AB}= X_{AB} s^{\epsilon} + Y_{AB} s^{-\eta}~~~~~~~~~~~~~~~~~~~\\
\epsilon=0.08, ~~~~\eta=0.45.
\eqno(1)
$$

For the 10 measured total cross-sections ($pp,~ \bar{p}p,~ pn,~
\bar{p}n, ~ \pi^{\pm} p, 
 ~ K^{\pm} p, ~ K^{\pm} n$) 15 (or 17) parameters (depending on whether
one requires $X_{np}=X_{pp}$ and $X_{K^+p}=X_{K^+n}$ or not)
are needed: five (or seven) 
$X$'s and ten $Y$'s. Taken at face value, these fits and 
subsequent refinements thereof \cite{DG} 
are at odds with a number of ideas grounded
in the quark model and a QCD-string approach to hadron scattering; we have
in mind ideas like
two-component duality, exchange degeneracy, Chan-Paton rules, 
flavor $U(3)$ symmetry, 
universality of vector-meson couplings.
We wish to show that one can 
successfully implement these ideas as constraints from the beginning 
and obtain fits different from, but 
of comparable quality to those of 
refs. \cite{DL, DG}.

\bigskip

{\bf 2. Some Theoretical Ideas on High Energy Hadron Scattering}

\bigskip

We start by explaining how each of the just mentioned ideas constrain
high energy hadronic total cross-sections.

{\em A) Universal and flavor U(3) symmetric vector meson coupling pattern}.

Although in QCD the nine light vector mesons appear as $q\bar{q}$
bound states whose coupling pattern is to be dynamically calculated,
it has been known for a long time that the observed 
pattern closely follows the pattern which
would be expected if these mesons were flavor gauge bosons.
Specifically this means that (suppressing Lorentz and Dirac indices) 
in the familiar $3\times 3$ matrix representation, the
coupling of these vector mesons $V$ to baryons $B$ is of the form
Tr$(\bar{B}[V,B]) +$Tr$(\bar{B} B)$Tr$V$ (here the ratio of the two terms' 
coefficients is determined by requiring the decoupling of the $\phi$ from the 
proton). Moreover 
$g_{\rho^{0} \bar{p} p}= \frac{1}{2} g_{\rho^{0} \pi^{-} \pi^{+}}$, since 
the third component of the proton's isospin is half that of the positive 
pion's. Strictly speaking, only at $t= m_{\rho}^2 \approx m_{\omega}^2$
does this coupling pattern 
determine 
the residue pattern of the 
odd-signature Regge poles on whose trajectory the vector mesons lie.
We will assume that the same pattern
is valid also at $t=0$. All this then yields four linear relations 
between the five odd-charge-conjugation total cross-section combinations.
With the notation $\Delta_{AB}=\sigma_{AB} - \sigma_{\bar{A} B}$,
where $\sigma_{AB}$ denotes the $AB$ total cross-section,
these four relations take the familiar form \cite{JT}, \cite{F}
$$
\Delta_{\bar{p} p}=5 \Delta_{\pi^{-} p} ~~~~~~~~
\Delta_{\bar{p} n}=4 \Delta_{\pi^{-} p}~~~~~~~~
\Delta_{K^{-} p}=2 \Delta_{\pi^{-} p}~~~~~~~
\Delta_{K^{-} n}=\Delta_{\pi^{-} p}.
\eqno(2)
$$
Of these five differences, $\Delta_{\pi^{-} p}$ involves only $\rho$ exchange,
whereas the remaining four involve both $\rho$ and $\omega$ exchange, 
predominantly the latter. The derivation of the relations (2) assumes the 
$\rho$ and $\omega$ trajectories' intercepts to be equal:
$\alpha_{\rho}(0)=\alpha_{\omega}(0)=1-\eta$. We first fit, in fig. 1a,
$\Delta_{\pi^{-} p}$ to an 
\vspace{-1cm}
\begin{figure}[h]
\begin{center}
\setlength{\unitlength}{0.1bp}
\subfigure{
\special{!
/gnudict 40 dict def
gnudict begin
/Color false def
/Solid false def
/gnulinewidth 5.000 def
/vshift -33 def
/dl {10 mul} def
/hpt 31.5 def
/vpt 31.5 def
/M {moveto} bind def
/L {lineto} bind def
/R {rmoveto} bind def
/V {rlineto} bind def
/vpt2 vpt 2 mul def
/hpt2 hpt 2 mul def
/Lshow { currentpoint stroke M
  0 vshift R show } def
/Rshow { currentpoint stroke M
  dup stringwidth pop neg vshift R show } def
/Cshow { currentpoint stroke M
  dup stringwidth pop -2 div vshift R show } def
/DL { Color {setrgbcolor Solid {pop []} if 0 setdash }
 {pop pop pop Solid {pop []} if 0 setdash} ifelse } def
/BL { stroke gnulinewidth 2 mul setlinewidth } def
/AL { stroke gnulinewidth 2 div setlinewidth } def
/PL { stroke gnulinewidth setlinewidth } def
/LTb { BL [] 0 0 0 DL } def
/LTa { AL [1 dl 2 dl] 0 setdash 0 0 0 setrgbcolor } def
/LT0 { PL [] 0 1 0 DL } def
/LT1 { PL [4 dl 2 dl] 0 0 1 DL } def
/LT2 { PL [2 dl 3 dl] 1 0 0 DL } def
/LT3 { PL [1 dl 1.5 dl] 1 0 1 DL } def
/LT4 { PL [5 dl 2 dl 1 dl 2 dl] 0 1 1 DL } def
/LT5 { PL [4 dl 3 dl 1 dl 3 dl] 1 1 0 DL } def
/LT6 { PL [2 dl 2 dl 2 dl 4 dl] 0 0 0 DL } def
/LT7 { PL [2 dl 2 dl 2 dl 2 dl 2 dl 4 dl] 1 0.3 0 DL } def
/LT8 { PL [2 dl 2 dl 2 dl 2 dl 2 dl 2 dl 2 dl 4 dl] 0.5 0.5 0.5 DL } def
/P { stroke [] 0 setdash
  currentlinewidth 2 div sub M
  0 currentlinewidth V stroke } def
/D { stroke [] 0 setdash 2 copy vpt add M
  hpt neg vpt neg V hpt vpt neg V
  hpt vpt V hpt neg vpt V closepath stroke
  P } def
/A { stroke [] 0 setdash vpt sub M 0 vpt2 V
  currentpoint stroke M
  hpt neg vpt neg R hpt2 0 V stroke
  } def
/B { stroke [] 0 setdash 2 copy exch hpt sub exch vpt add M
  0 vpt2 neg V hpt2 0 V 0 vpt2 V
  hpt2 neg 0 V closepath stroke
  P } def
/C { stroke [] 0 setdash exch hpt sub exch vpt add M
  hpt2 vpt2 neg V currentpoint stroke M
  hpt2 neg 0 R hpt2 vpt2 V stroke } def
/T { stroke [] 0 setdash 2 copy vpt 1.12 mul add M
  hpt neg vpt -1.62 mul V
  hpt 2 mul 0 V
  hpt neg vpt 1.62 mul V closepath stroke
  P  } def
/S { 2 copy A C} def
end
}
\begin{picture}(2700,2160)(0,300)
\special{"
gnudict begin
gsave
50 50 translate
0.100 0.100 scale
0 setgray
/Helvetica findfont 100 scalefont setfont
newpath
-500.000000 -500.000000 translate
LTa
LTb
600 251 M
63 0 V
1854 0 R
-63 0 V
600 561 M
63 0 V
1854 0 R
-63 0 V
600 870 M
63 0 V
1854 0 R
-63 0 V
600 1180 M
63 0 V
1854 0 R
-63 0 V
600 1490 M
63 0 V
1854 0 R
-63 0 V
600 1799 M
63 0 V
1854 0 R
-63 0 V
600 2109 M
63 0 V
1854 0 R
-63 0 V
826 251 M
0 63 V
0 1795 R
0 -63 V
1107 251 M
0 63 V
0 1795 R
0 -63 V
1389 251 M
0 63 V
0 1795 R
0 -63 V
1671 251 M
0 63 V
0 1795 R
0 -63 V
1953 251 M
0 63 V
0 1795 R
0 -63 V
2235 251 M
0 63 V
0 1795 R
0 -63 V
2517 251 M
0 63 V
0 1795 R
0 -63 V
600 251 M
1917 0 V
0 1858 V
-1917 0 V
600 251 L
LT0
2214 1946 M
180 0 V
600 1665 M
19 -117 V
20 -78 V
19 -57 V
19 -44 V
20 -35 V
19 -29 V
20 -24 V
19 -21 V
19 -18 V
20 -16 V
19 -14 V
19 -12 V
20 -12 V
19 -10 V
19 -9 V
20 -9 V
19 -8 V
20 -7 V
19 -7 V
19 -7 V
20 -5 V
19 -6 V
19 -5 V
20 -5 V
19 -5 V
19 -4 V
20 -5 V
19 -4 V
20 -3 V
19 -4 V
19 -3 V
20 -4 V
19 -3 V
19 -3 V
20 -3 V
19 -2 V
19 -3 V
20 -3 V
19 -2 V
20 -3 V
19 -2 V
19 -2 V
20 -2 V
19 -2 V
19 -2 V
20 -2 V
19 -2 V
19 -2 V
20 -2 V
19 -1 V
20 -2 V
19 -2 V
19 -1 V
20 -2 V
19 -1 V
19 -2 V
20 -1 V
19 -2 V
19 -1 V
20 -1 V
19 -2 V
20 -1 V
19 -1 V
19 -1 V
20 -1 V
19 -2 V
19 -1 V
20 -1 V
19 -1 V
19 -1 V
20 -1 V
19 -1 V
20 -1 V
19 -1 V
19 -1 V
20 -1 V
19 -1 V
19 -1 V
20 0 V
19 -1 V
19 -1 V
20 -1 V
19 -1 V
20 -1 V
19 0 V
19 -1 V
20 -1 V
19 -1 V
19 0 V
20 -1 V
19 -1 V
19 0 V
20 -1 V
19 -1 V
20 0 V
19 -1 V
19 -1 V
20 0 V
19 -1 V
LT1
610 1492 D
620 1498 D
621 1405 D
626 1421 D
630 1730 D
631 1584 D
636 1211 D
644 1356 D
652 1408 D
653 1310 D
663 1379 D
668 1273 D
678 1291 D
705 1304 D
731 1245 D
758 1296 D
784 1220 D
811 1247 D
837 1285 D
864 1188 D
916 1112 D
1075 1078 D
1181 1069 D
1324 1025 D
1340 1059 D
1445 1034 D
1604 1040 D
1816 1025 D
2027 966 D
2186 994 D
2345 1013 D
610 1405 M
0 174 V
579 1405 M
62 0 V
-62 174 R
62 0 V
620 901 M
0 1194 V
589 901 M
62 0 V
589 2095 M
62 0 V
621 1370 M
0 71 V
-31 -71 R
62 0 V
-62 71 R
62 0 V
626 1272 M
0 299 V
595 1272 M
62 0 V
-62 299 R
62 0 V
630 1388 M
0 683 V
599 1388 M
62 0 V
-62 683 R
62 0 V
631 1222 M
0 724 V
600 1222 M
62 0 V
-62 724 R
62 0 V
636 352 M
0 1718 V
605 352 M
62 0 V
605 2070 M
62 0 V
644 1151 M
0 410 V
613 1151 M
62 0 V
-62 410 R
62 0 V
652 1274 M
0 269 V
621 1274 M
62 0 V
-62 269 R
62 0 V
653 1274 M
0 71 V
-31 -71 R
62 0 V
-62 71 R
62 0 V
-21 -5 R
0 78 V
-31 -78 R
62 0 V
-62 78 R
62 0 V
668 1181 M
0 183 V
637 1181 M
62 0 V
-62 183 R
62 0 V
678 1089 M
0 405 V
647 1089 M
62 0 V
-62 405 R
62 0 V
-4 -359 R
0 338 V
674 1135 M
62 0 V
-62 338 R
62 0 V
-5 -315 R
0 174 V
700 1158 M
62 0 V
-62 174 R
62 0 V
-4 -202 R
0 331 V
727 1130 M
62 0 V
-62 331 R
62 0 V
-5 -444 R
0 406 V
753 1017 M
62 0 V
-62 406 R
62 0 V
-4 -248 R
0 144 V
780 1175 M
62 0 V
-62 144 R
62 0 V
-5 -237 R
0 406 V
806 1082 M
62 0 V
-62 406 R
62 0 V
-4 -474 R
0 347 V
833 1014 M
62 0 V
-62 347 R
62 0 V
21 -341 R
0 183 V
885 1020 M
62 0 V
-62 183 R
62 0 V
1075 986 M
0 183 V
1044 986 M
62 0 V
-62 183 R
62 0 V
75 -190 R
0 179 V
1150 979 M
62 0 V
-62 179 R
62 0 V
1324 857 M
0 337 V
1293 857 M
62 0 V
-62 337 R
62 0 V
1340 970 M
0 179 V
1309 970 M
62 0 V
-62 179 R
62 0 V
74 -204 R
0 179 V
1414 945 M
62 0 V
-62 179 R
62 0 V
128 -104 R
0 41 V
-31 -41 R
62 0 V
-62 41 R
62 0 V
181 -63 R
0 55 V
-31 -55 R
62 0 V
-62 55 R
62 0 V
2027 930 M
0 72 V
-31 -72 R
62 0 V
-62 72 R
62 0 V
128 -47 R
0 79 V
-31 -79 R
62 0 V
-62 79 R
62 0 V
128 -71 R
0 100 V
2314 963 M
62 0 V
-62 100 R
62 0 V
stroke
grestore
end
}
\put(2154,1946){\makebox(0,0)[r]{12.93s$^{-0.5397}$}}
\put(1558,-50){\makebox(0,0){s \small{(GeV$^2$)}}}
\put(1548,-220){\makebox(0,0){\footnotesize (a)}}
\put(350,1180){%
\special{ps: gsave currentpoint currentpoint translate
270 rotate neg exch neg exch translate}%
\makebox(0,0)[b]{\shortstack{$\Delta_{\pi^-p}$ \small{(mb)}}}%
\special{ps: currentpoint grestore moveto}%
}
\put(2517,151){\makebox(0,0){700}}
\put(2235,151){\makebox(0,0){600}}
\put(1953,151){\makebox(0,0){500}}
\put(1671,151){\makebox(0,0){400}}
\put(1389,151){\makebox(0,0){300}}
\put(1107,151){\makebox(0,0){200}}
\put(826,151){\makebox(0,0){100}}
\put(540,2109){\makebox(0,0)[r]{4}}
\put(540,1799){\makebox(0,0)[r]{3}}
\put(540,1490){\makebox(0,0)[r]{2}}
\put(540,1180){\makebox(0,0)[r]{1}}
\put(540,870){\makebox(0,0)[r]{0}}
\put(540,561){\makebox(0,0)[r]{-1}}
\put(540,251){\makebox(0,0)[r]{-2}}
\end{picture}
}
\end{center}
\end{figure}
\begin{figure}[p,b,t]
\input{inc.tex}
\caption{Single Regge pole fits constrained by Eqs. (2) \ to the odd signature cross section differences.}
\end{figure}

\clearpage
\noindent expression of the form
$$
\Delta_{\pi^{-} p}=\delta_{\pi p} s^{-\eta}
\eqno(3a)
$$
and obtain
$$
\eta=0.54 ~~~~~~~~~ \delta_{\pi p} =12.93
\eqno(3b)
$$

Excellent fits to the remaining four cross-section differences are then 
obtained by multiplying the function (3) by the integers given in Eqs. (2).
We have checked that, not surprisingly, the parameters obtained in the 
fits of Refs. \cite{DL}, \cite{DG} also obey the constraints imposed
upon them by Eqs. (2).

We should point out that 
here and throughout 
this paper we normalize coefficients so that $s$ is measured in GeV$^2$ and 
we use the data of ref. \cite{DG}.

{\em B) Exchange Degeneracy/ Chan-Paton Rules.}

In the limit of large number of colors, QCD reduces to a string theory in 
which mesons are open strings with a quark at one end and an antiquark
at the other (fig. 2a). The strings themselves are tubes of color-electric 
flux.
Baryons are also viewed as systems of three strings with a quark at each
of the three open ends, the three other ends meeting at a node (fig. 2b).
\begin{figure}[b,t]
\input{strings.tex}
\caption{Mesonic and baryonic strings (\mbox{$\times=\bar{q}, \circ=q$})}
\end{figure}
When this string picture applies, hopefully for 3 colors already, then 
hadronic amplitudes obey duality (we use this word in the sense that the
sum of the resonance contributions in the $s$-channel gives 
rise to the imaginary part of the Regge contribution in the crossed
$t$-channel). This, in turn, requires the degeneracy of the odd and even
$t$-channel mesonic
Regge pole trajectories and the equality of their residues, for otherwise
the amplitude for say $pp$ scattering would have a Regge pole 
contribution with nonvanishing imaginary part even though there are no
$pp$ resonances. The near degeneracy of the 
observed $\omega$ and $\rho$ meson masses on the one hand and of 
the $f$ and $a2$ meson masses on the other is as required by exchange
degeneracy. The equality of the isospin $I=1$ 
residues yields further constraints, to wit
$$
\sigma_{pn} -\sigma_{pp} = 0 ~~~~~~~~~~~~~~~~~~~~
\sigma_{K^{+} n} -\sigma_{K^{+} p} = 0
\eqno(4)
$$
As can be seen from fig. 3, the kaon difference is compatible with zero, 
but the $pn-pp$ difference, though very small, appears not to strictly vanish.
It can be fit to a combination of a Pomeron-Regge cut and of a small
Regge pole term, though the individual contribution of these terms is
hard to determine from the data. Even a pure cut gives a good fit.
We can therefore safely assume that at string tree level both equations (4) are
obeyed.
By combining 
with the just discussed $I=1$ exchange degeneracy relations
their isospin $I=0$ counterparts,
one imposes the full Chan-Paton rules and this then requires the absence 
of a mesonic Regge exchange contribution in $pp, pn, K^{+}p$ and $K^{+} n$
total cross-sections. This requirement is strongly violated in the fits
of references \cite{DL, DG}. The reason for this is simple to understand.
Before its ultimate rise at very high energies,\begin{figure}[h]
\setlength{\unitlength}{0.1bp}
{\mbox{\hspace*{-3cm} \subfigure[]{
\begin{picture}(2700,2160)(0,0)
\put(1558,-50){\makebox(0,0){s \small{(GeV$^2$)}}}
\put(350,1180){%
\makebox(0,0)[b]{\shortstack{$\sigma_{pn}-\sigma_{pp}$ (mb)}}%
}
\put(2517,151){\makebox(0,0){700}}
\put(2235,151){\makebox(0,0){600}}
\put(1953,151){\makebox(0,0){500}}
\put(1671,151){\makebox(0,0){400}}
\put(1389,151){\makebox(0,0){300}}
\put(1107,151){\makebox(0,0){200}}
\put(826,151){\makebox(0,0){100}}
\put(540,2109){\makebox(0,0)[r]{4}}
\put(540,1844){\makebox(0,0)[r]{3}}
\put(540,1578){\makebox(0,0)[r]{2}}
\put(540,1313){\makebox(0,0)[r]{1}}
\put(540,1047){\makebox(0,0)[r]{0}}
\put(540,782){\makebox(0,0)[r]{-1}}
\put(540,516){\makebox(0,0)[r]{-2}}
\put(540,251){\makebox(0,0)[r]{-3}}
\end{picture}
}\hspace{-1cm}
\subfigure[]{
\begin{picture}(2700,2160)(0,0)
\put(1558,-50){\makebox(0,0){s \small{(GeV$^2$)}}}
\put(300,1180){%
\makebox(0,0)[b]{\shortstack{$\sigma_{K^+p}-\sigma_{K^+n}$ \small{(mb)}}}%
}
\put(2517,151){\makebox(0,0){600}}
\put(2186,151){\makebox(0,0){500}}
\put(1856,151){\makebox(0,0){400}}
\put(1525,151){\makebox(0,0){300}}
\put(1195,151){\makebox(0,0){200}}
\put(864,151){\makebox(0,0){100}}
\put(540,2109){\makebox(0,0)[r]{2}}
\put(540,1578){\makebox(0,0)[r]{1}}
\put(540,1047){\makebox(0,0)[r]{0}}
\put(540,516){\makebox(0,0)[r]{-1}}
\end{picture}}}}
\caption{Test of the I=1 exchange degeneracy relations (4).\
The solid curve in (a) represents a pure Pomeron-Regge cut fit.}
\end{figure}
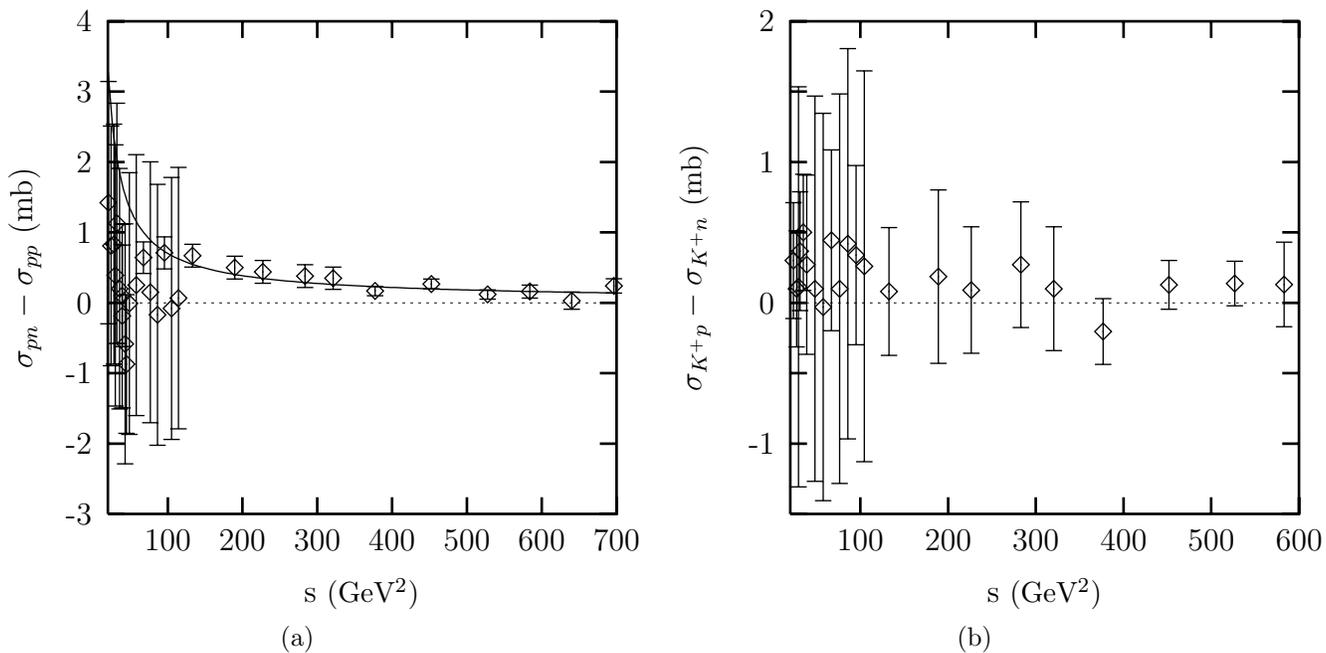
 
any of these cross-sections, $\sigma_{pp}$ in 
particular, decreases with energy. In a fit of type (1) this is only 
possible if a Regge term is present and the particular Regge term needed
to fit the decrease in the low energy $\sigma_{pp}$ turns out to be very large:
$\approx 7\Delta_{\pi^{-} p}$. The fits of type (1) make two simplifying,
but otherwise arbitrary
assumptions, one concerning the nature of the Pomeron 
as a unique ``effective" Regge pole, and the other
concerning the absence of Regge-Regge cuts.
We shall see below that by relaxing these assumptions, 
fits of comparable quality
which do not violate this $I=0$ exchange degeneracy requirement are readily 
obtained. 

\bigskip
{\bf 3. Experimental Test of Principles A) and B)}
\bigskip

Before we get to these new fits, we must first analyse in some detail the full 
implications of the assumptions A) and B) above. There are 10 
measured total cross-sections and Eqs. (2) and (4) provide 6 linear relations 
among them, thus leaving 4 independent combinations, which we choose as
$\Delta_{\pi^{-} p}$, $\sigma_{pp}$, $\sigma_{\pi^{+} p}$ and 
$\sigma_{ K^{+} p}$. Of these, the odd charge-conjugation combination
$\Delta_{\pi^{-} p}$ is dominated, 
as was already mentioned, by the exchange of the $\rho$ Regge pole
and this is well borne out by the data, as was known for decades. So we
really have to fit only the remaining three cross-sections. To do so, let us
consider each of them separately. Let us start with 
$\sigma_{pp}$. We write for it the generic formula
$$
 \sigma_{pp}= P_p(s) + Y_p s^{-\eta} +Z_p s^{-\lambda},
\eqno(5)
$$
where $P_p(s)$ is the Pomeron contribution, $Y_p s^{-\eta}$ is the 
$f$-$\omega$-$a2$-$\rho$ Regge
contribution and $Z_p s^{-\lambda}$ is a contribution due to Regge-Regge
cuts and to the $f'$ Regge pole. Now let us consider each of these terms.
First of all, the Regge contribution $Y_p s^{-\eta}$ would be absent at
the string tree level.
At this level the other two terms would be absent as well. Indeed in
a string approach, the Pomeron is ``$f$-dominated" \cite{CGZ, CF} 
at both ends. In other words,
one of the protons emits an open $f$-string, which closes up into a Pomeron
and then reopens into another $f$-string which gets 
absorbed by the other proton (see fig. 4).
This process gets iterated and at the next and later 
steps involves the $f'$ as well, as shown in fig. 4. The consecutive steps 
are suppressed by OZI rule breaking so that the ensuing breaking of
exchange degeneracy is small. There is strong evidence in favor of this 
``$f$-dominated" Pomeron in the photoproduction of the $\rho, \omega,
\phi, J/\psi$ vector mesons \cite{CF}. For us
the important point is that {\em exchange 
degeneracy is exact  only at the string tree level}. 
Its small breaking is caused 
primarily by 
Pomeron-$f$-$f'$ mixing. As such, the first two terms in 
Eq. (5) are expected to be there, 
with the understanding 
\newpage
\begin{figure}[hp]
\vspace*{-1cm}
\setlength{\unitlength}{0.012500in}%
\begingroup\makeatletter
\def\x#1#2#3#4#5#6#7\relax{\def\x{#1#2#3#4#5#6}}%
\expandafter\x\fmtname xxxxxx\relax \def\y{splain}%
\ifx\x\y   
\gdef\SetFigFont#1#2#3{%
  \ifnum #1<17\tiny\else \ifnum #1<20\small\else
  \ifnum #1<24\normalsize\else \ifnum #1<29\large\else
  \ifnum #1<34\Large\else \ifnum #1<41\LARGE\else
     \huge\fi\fi\fi\fi\fi\fi
  \csname #3\endcsname}%
\else
\gdef\SetFigFont#1#2#3{\begingroup
  \count@#1\relax \ifnum 25<\count@\count@25\fi
  \def\x{\endgroup\@setsize\SetFigFont{#2pt}}%
  \expandafter\x
    \csname \romannumeral\the\count@ pt\expandafter\endcsname
    \csname @\romannumeral\the\count@ pt\endcsname
  \csname #3\endcsname}%
\fi
\endgroup
\begin{picture}(384,460)(80,340)
\thicklines
\put(185,658){\line( 0,-1){ 76}}
\put(185,620){\line( 1, 0){ 34}}
\put(354,658){\line( 0,-1){ 76}}
\put(219,620){\linethickness{9pt}\line( 1, 0){ 34}}
\put(287,620){\linethickness{9pt}\line( 1, 0){ 34}}
\put(359,578){\makebox(0,0)[lb]{\smash{\SetFigFont{10}{12.0}{rm}p}}}
\put(177,578){\makebox(0,0)[lb]{\smash{\SetFigFont{10}{12.0}{rm}p}}}
\put(202,603){\makebox(0,0)[lb]{\smash{\SetFigFont{10}{12.0}{rm}f}}}
\put(236,603){\makebox(0,0)[lb]{\smash{\SetFigFont{10}{12.0}{rm}P}}}
\put(304,603){\makebox(0,0)[lb]{\smash{\SetFigFont{10}{12.0}{rm}P}}}
\put(261,603){\makebox(0,0)[lb]{\smash{\SetFigFont{10}{12.0}{rm}f,f$'$}}}
\put(337,603){\makebox(0,0)[lb]{\smash{\SetFigFont{10}{12.0}{rm}f}}}
\put(371,616){\makebox(0,0)[lb]{\smash{\SetFigFont{14}{16.8}{rm}+}}}
\put(388,620){\makebox(0,0)[lb]{\smash{\SetFigFont{14}{16.8}{rm}.  .  .  .}}}
\put(177,658){\makebox(0,0)[lb]{\smash{\SetFigFont{10}{12.0}{rm}p}}}
\put(359,658){\makebox(0,0)[lb]{\smash{\SetFigFont{10}{12.0}{rm}p}}}
\put(253,620){\line( 1, 0){ 34}}
\put(321,620){\makebox(0.4444,0.6667){\SetFigFont{10}{12}{rm}.}}
\put(321,620){\line( 1, 0){ 33}}
\put(354,620){\line( 0, 1){  0}}
\put(354,620){\line( 0, 1){  0}}
\put(329,754){\linethickness{9pt}\line( 1, 0){ 59}}
\put(278,792){\line( 0,-1){ 76}}
\put(278,716){\line( 0, 1){  5}}
\put(278,754){\line( 1, 0){ 64}}
\put(434,792){\line( 0,-1){ 76}}
\put(439,792){\makebox(0,0)[lb]{\smash{\SetFigFont{10}{12.0}{rm}p}}}
\put(299,738){\makebox(0,0)[lb]{\smash{\SetFigFont{10}{12.0}{rm}f}}}
\put(354,738){\makebox(0,0)[lb]{\smash{\SetFigFont{10}{12.0}{rm}P}}}
\put(413,738){\makebox(0,0)[lb]{\smash{\SetFigFont{10}{12.0}{rm}f}}}
\put(270,712){\makebox(0,0)[lb]{\smash{\SetFigFont{10}{12.0}{rm}p}}}
\put(456,750){\makebox(0,0)[lb]{\smash{\SetFigFont{14}{16.8}{rm}+}}}
\put(270,792){\makebox(0,0)[lb]{\smash{\SetFigFont{10}{12.0}{rm}p}}}
\put(321,754){\line( 1, 0){113}}
\put(439,712){\makebox(0,0)[lb]{\smash{\SetFigFont{10}{12.0}{rm}p}}}
\put( 88,792){\line( 0, 1){  0}}
\put( 88,792){\line( 0,-1){ 76}}
\put(190,792){\line( 0,-1){ 76}}
\put( 88,754){\line( 1, 0){102}}
\put( 80,792){\makebox(0,0)[lb]{\smash{\SetFigFont{10}{12.0}{rm}p}}}
\put(194,792){\makebox(0,0)[lb]{\smash{\SetFigFont{10}{12.0}{rm}p}}}
\put( 80,712){\makebox(0,0)[lb]{\smash{\SetFigFont{10}{12.0}{rm}p}}}
\put(194,712){\makebox(0,0)[lb]{\smash{\SetFigFont{10}{12.0}{rm}p}}}
\put(232,750){\makebox(0,0)[lb]{\smash{\SetFigFont{14}{16.8}{rm}+}}}
\put(139,616){\makebox(0,0)[lb]{\smash{\SetFigFont{14}{16.8}{rm}+}}}
\put(139,738){\makebox(0,0)[lb]{\smash{\SetFigFont{10}{12.0}{rm}f}}}
\put(180,448){\line( 0, 1){  0}}
\multiput(180,448)(0.36364,0.45455){12}{\makebox(0.4444,0.6667){\SetFigFont{7}{8.4}{rm}.}}
\multiput(180,453)(0.36364,-0.45455){12}{\makebox(0.4444,0.6667){\SetFigFont{7}{8.4}{rm}.}}
\put(182,416){\circle{6}}
\put(182,451){\line( 0,-1){ 32}}
\put(148,373){\circle{6}}
\put(122,396){\line( 1, 0){  6}}
\put(126,398){\line( 0,-1){  6}}
\put(126,396){\line( 1,-1){ 21}}
\put(157,381){\circle{6}}
\put(131,404){\line( 1, 0){  6}}
\put(134,407){\line( 0,-1){  6}}
\put(134,404){\line( 1,-1){ 20.500}}
\put(165,389){\circle{6}}
\put(139,413){\line( 1, 0){  6}}
\put(142,415){\line( 0,-1){  6}}
\put(142,413){\line( 1,-1){ 22}}
\put(140,364){\circle{6}}
\put(114,387){\line( 1, 0){  6}}
\put(117,390){\line( 0,-1){  5}}
\put(117,387){\line( 1,-1){ 20.500}}
\put(164,474){\circle{6}}
\put(139,451){\line( 1, 0){  6}}
\put(142,455){\line( 0,-1){  6}}
\put(142,451){\line( 1, 1){ 20.500}}
\put(156,482){\circle{6}}
\put(131,460){\line( 1, 0){  6}}
\put(133,463){\line( 0,-1){  6}}
\put(133,460){\line( 1, 1){ 19.500}}
\put(148,490){\circle{6}}
\put(122,468){\line( 1, 0){  6}}
\put(125,472){\line( 0,-1){  6}}
\put(125,468){\line( 1, 1){ 20}}
\put(139,499){\circle{6}}
\put(114,477){\line( 1, 0){  6}}
\put(116,479){\line( 0,-1){  5}}
\put(116,477){\line( 6, 5){ 21.738}}
\put(413,386){\circle{6}}
\put(388,364){\line( 1, 0){  6}}
\put(391,367){\line( 0,-1){  6}}
\put(391,364){\line( 1, 1){ 20}}
\put(405,394){\circle{6}}
\put(380,372){\line( 1, 0){  6}}
\put(382,375){\line( 0,-1){  5}}
\put(382,372){\line( 1, 1){ 19.500}}
\put(396,402){\circle{6}}
\put(371,381){\line( 1, 0){  6}}
\put(374,384){\line( 0,-1){  6}}
\put(374,381){\line( 1, 1){ 19.500}}
\put(388,411){\circle{6}}
\put(363,388){\line( 1, 0){  6}}
\put(365,391){\line( 0,-1){  5}}
\put(365,388){\line( 1, 1){ 20.500}}
\put(397,461){\circle{6}}
\put(371,483){\line( 1, 0){  6}}
\put(375,486){\line( 0,-1){  6}}
\put(375,483){\line( 1,-1){ 20.500}}
\put(406,469){\circle{6}}
\put(380,492){\line( 1, 0){  6}}
\put(383,494){\line( 0,-1){  5}}
\put(383,492){\line( 1,-1){ 20.500}}
\put(414,478){\circle{6}}
\put(388,500){\line( 1, 0){  6}}
\put(391,503){\line( 0,-1){  6}}
\put(391,500){\line( 1,-1){ 21.500}}
\put(389,452){\circle{6}}
\put(363,476){\line( 1, 0){  6}}
\put(366,478){\line( 0,-1){  5}}
\put(366,476){\line( 1,-1){ 21}}
\put(349,448){\line( 0, 1){  0}}
\multiput(349,448)(0.36364,0.45455){12}{\makebox(0.4444,0.6667){\SetFigFont{7}{8.4}{rm}.}}
\multiput(349,453)(0.36364,-0.45455){12}{\makebox(0.4444,0.6667){\SetFigFont{7}{8.4}{rm}.}}
\put(351,416){\circle{6}}
\put(351,451){\line( 0,-1){ 32}}
\multiput(198,449)(0.40000,0.40000){11}{\makebox(0.4444,0.6667){\SetFigFont{7}{8.4}{rm}.}}
\multiput(198,453)(0.40000,-0.40000){11}{\makebox(0.4444,0.6667){\SetFigFont{7}{8.4}{rm}.}}
\put(200,435){\oval(  8, 32)[tl]}
\put(200,435){\oval(  8, 34)[bl]}
\put(200,415){\circle{6}}
\multiput(334,449)(-0.40000,0.40000){11}{\makebox(0.4444,0.6667){\SetFigFont{7}{8.4}{rm}.}}
\multiput(334,453)(-0.40000,-0.40000){11}{\makebox(0.4444,0.6667){\SetFigFont{7}{8.4}{rm}.}}
\put(332,435){\oval(  8, 34)[br]}
\put(332,435){\oval(  8, 32)[tr]}
\put(332,415){\circle{6}}
\put(228,433){\oval( 36, 36)[tl]}
\put(228,433){\oval( 36, 38)[bl]}
\multiput(225,448)(0.42857,0.35714){15}{\makebox(0.4444,0.6667){\SetFigFont{7}{8.4}{rm}.}}
\multiput(225,453)(0.42857,-0.35714){15}{\makebox(0.4444,0.6667){\SetFigFont{7}{8.4}{rm}.}}
\put(232,414){\circle{6}}
\put(303,433){\oval( 34, 36)[br]}
\put(303,433){\oval( 34, 36)[tr]}
\multiput(305,448)(-0.42857,0.35714){15}{\makebox(0.4444,0.6667){\SetFigFont{7}{8.4}{rm}.}}
\multiput(305,453)(-0.42857,-0.35714){15}{\makebox(0.4444,0.6667){\SetFigFont{7}{8.4}{rm}.}}
\put(299,415){\circle{6}}
\put(266,432){\circle{42}}
\put(198,386){\makebox(0,0)[lb]{\smash{\SetFigFont{10}{12.0}{rm}f}}}
\put(333,386){\makebox(0,0)[lb]{\smash{\SetFigFont{10}{12.0}{rm}f}}}
\put(266,386){\makebox(0,0)[lb]{\smash{\SetFigFont{10}{12.0}{rm}P}}}
\put(114,486){\makebox(0,0)[lb]{\smash{\SetFigFont{10}{12.0}{rm}$\pi^+$}}}
\put(114,365){\makebox(0,0)[lb]{\smash{\SetFigFont{10}{12.0}{rm}$\pi^+$}}}
\put(418,486){\makebox(0,0)[lb]{\smash{\SetFigFont{10}{12.0}{rm}$\pi^+$}}}
\put(418,365){\makebox(0,0)[lb]{\smash{\SetFigFont{10}{12.0}{rm}$\pi^+$}}}
\put(257,558){\makebox(0,0)[lb]{\smash{\SetFigFont{10}{12.0}{rm}(a)}}}
\put(257,340){\makebox(0,0)[lb]{\smash{\SetFigFont{10}{12.0}{rm}(b)}}}
\end{picture}
\caption{The $f$-dominated Pomeron}
\end{figure}
\noindent that the coefficient of the Regge term is small when compared to 
that in the fit to $\Delta_{\pi^- p}$. The last term in Eq. (5) represents 
the contribution of Regge-Regge cuts and of the $f'$ Regge pole term induced 
by the string loop effect of  Pomeron-$f$-$f'$ mixing (see fig.4). 
Both the Regge-Regge cuts and the $f'$ pole have an intercept $\sim 0$,
so  $\lambda\approx 1$ in Eq. (5).

We now turn to the Pomeron term $P_p(s)$. This term is a stand-in for the 
Pomeron Regge pole and for the multi-Pomeron cuts, as was already
pointed out in ref. \cite{DL}. There all this complexity was lumped
into a unique power law $s^\epsilon$, for reasons of simplicity,
rather than on the basis of any theoretical considerations. 
Here we will relax this ``simplicity" constraint and set

$$
P_p(s) = X_p s^\epsilon + C_p s^\mu  ~~~~~~~~~~~~ 0\leq \mu<\epsilon
\eqno(6)
$$
The best fits we obtain for $\mu=0$, so that the last term in Eq. (6)
will be a constant.

We will thus simultaneously fit $\sigma_{pp}$, $\sigma_{\pi^+p}$ and 
$\sigma_{K^+p}$ to the forms:
\setcounter{equation}{6}
\begin{eqnarray}
\sigma_{pp}    & = & X_p s^\epsilon + C_p  + Y_p s^{-\eta} +Z_p s^{-\lambda}
\nonumber\\
\sigma_{\pi^+p} & = &X_{\pi} s^\epsilon + C_{\pi}+ (Y_{\pi}+\delta_{\pi p}) s^{-\eta} +Z_{\pi} s^{-\lambda}\\
\sigma_{K^+p}   & = & X_K s^\epsilon + C_K 
 + Y_K s^{-\eta} +Z_K s^{-\lambda} \nonumber
\end{eqnarray}

The two old parameters $\eta$ and $\delta_{\pi p}$,
which appear here, have been
determined above from fitting the odd charge conjugation combinations of
total cross-sections: $\eta=0.54$, $\delta_{\pi p}=12.93$. That it is
precisely $\delta_{\pi p}$ which appears in Eq. (7) is a straightforward 
consequence of assumptions A) and B) above.
The 14 new parameters which appear in Eq. (7) are fortunately 
not all independent or unconstrained.
First, all the $Y$'s originate in exchange degeneracy
breaking and must therefore be small compared to $\delta_{\pi p}$. The value 
of $\lambda$ must, as we saw, be near $1$. The quark model determines
the ratios
$$
\frac{3X_{\pi}}{2X_p} \approx \frac{3C_{\pi}}{2C_p} \approx 1
\eqno(8a)
$$
and the ``$f$-dominated Pomeron" requires 
$$
\frac{X_K}{X_{\pi}} \approx \frac{C_K}{C_{\pi}} \approx 
\frac{1}{2}\Bigl(1+\frac{m_{\rho}^2}{m_{\phi}^2} \Bigr)=0.886.
\eqno(8b)
$$
Eq. (7) then really introduces only 7 parameters and the small departures
from unity of the other just mentioned combinations of parameters. 
With all this in mind we now present our fits of type Eq. (7) in fig. 5.
The corresponding values of the parameters are
\setcounter{equation}{8}
\begin{eqnarray}
\epsilon=0.135 ~~~~~ \eta&=&0.54 ~~~~~ \lambda=1.01\nonumber \\
X_{p}=6.26 ~~~~~C_{p}=24.4 ~~&~&~~Y_p=0.88 ~~~~ Z_p=196 \nonumber \\
 \frac{3X_{\pi}}{2X_p} = 1.04 ~~~~ \frac{3C_{\pi}}{2C_p}=0.84 ~~&~&~~ Y_{\pi}=-1.9~~~~~ Z_{\pi}=51 \\
\frac{X_K}{X_{\pi}} = 0.9 ~~~~ \frac{C_K}{C_{\pi}} = 0.83 ~~&~&~~ Y_K=-0.88~~~~ Z_K=0.12.\nonumber
\end{eqnarray}
\begin{figure}[p]
\vspace*{-1cm}
\subfigure{\begin{picture}(450,500)(50,50)
\put(0,0){
\epsfig{file=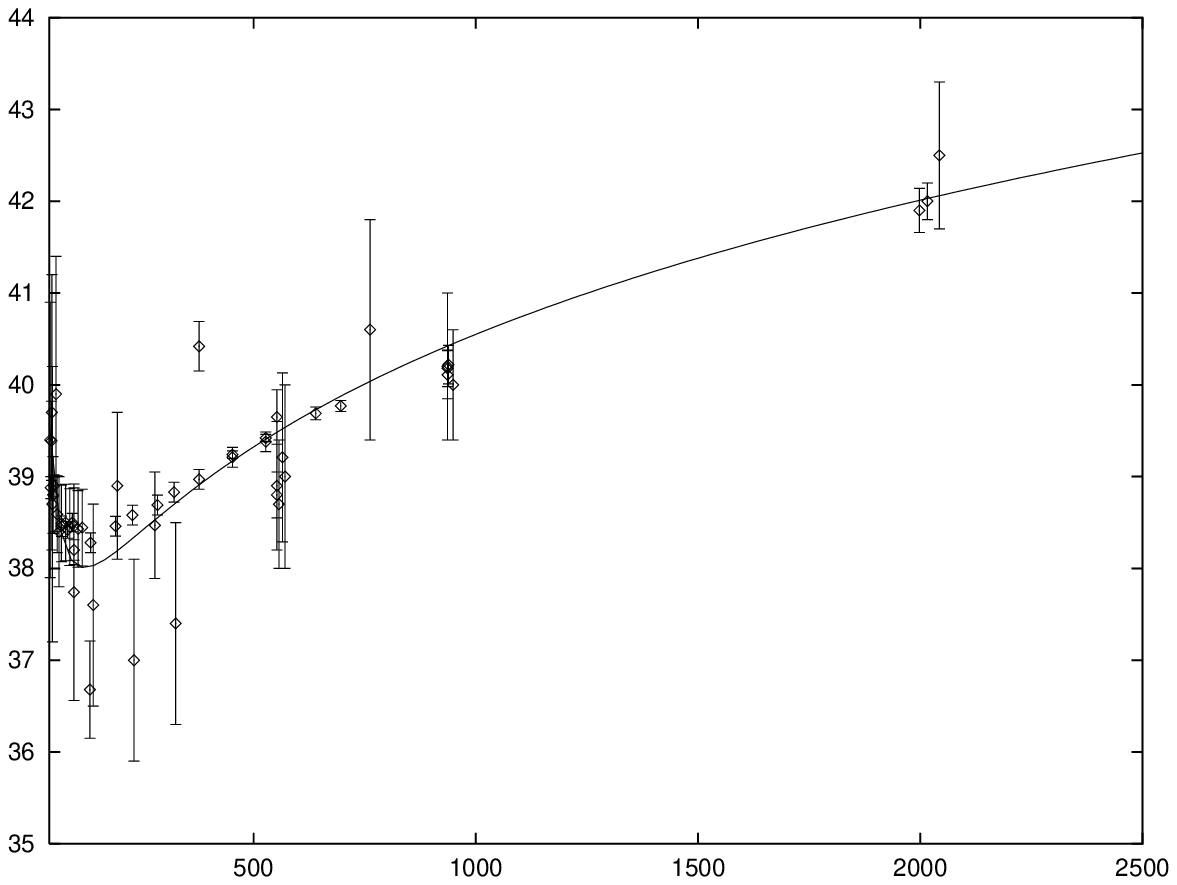, width = 8in, height=6in, angle=90}
}
\put(200,15){\makebox(0,0){$\sigma_{pp}$ \small{(mb)}}}
\put(460,300){%
\makebox(0,0)[b]{\shortstack{s \small{(GeV$^2$)}}}%
}
\put(475,300){%
\makebox(0,0)[b]{\shortstack{\footnotesize (a)}}%
}
\end{picture}}

\end{figure}
\input{inc3.tex}
This fit is of comparable quality to the fits of refs. \cite{DL}, \cite{DG}.
It has the following characteristic features: 

--- The exchange degeneracy breaking parameters $Y_p$, $Y_{\pi}$ and 
$Y_K$ are indeed very small as compared to $\Delta_{\pi p}$.

---The Pomeron exponent $\epsilon$ is larger than in most previous fits,
but such a larger value was already contemplated in ref. \cite{A} in the 
context of very high energies. In any case, even with this larger exponent 
the Froissart bound is comfortably obeyed even beyond the Planck energy.

--- The constraints (8) are well obeyed by the leading Pomeron terms ($X$
coefficients) and obeyed at the 15\% level for the subdominant terms 
($C$ coefficients).

--- At first sight the $f'$-Regge-Regge-cut parameter $Z_p$ appears 
large. $Z_p$ plays for our fit a role similar to that of the large exchange
degeneracy breaking $Y_p$ parameter in refs. \cite{DL}, \cite{DG}. This 
$f'$-Regge-Regge-cut term falls much faster with energy than an 
ordinary Regge term, so it makes sense to compare its low energy contribution
in the $pp$ amplitude, ${\cal CUT} \sim 196s^{-\lambda+1}\approx 196$,
to the {\em nonvanishing real part} of the $pp$ Regge term which is 
Re$(R) \sim 5\Delta_{\pi p} s^{-\eta+ 1} \approx 65 s^{0.46}$. Even at the low
value $s=40$ GeV$^2$, we find ${\cal CUT} /$Re$(R) \approx 0.55$ 
and this ratio 
decreases as $s^{-0.46}$. The Regge cut and $f'$ contributions
are thus consistently smaller than the Regge pole contributions.
Similar arguments can be made for the $\pi p$ and $Kp$ amplitudes as well.
The important new feature here is that this $f'$-Regge-Regge-cut term 
represents a theoretically expected exchange degeneracy and duality violation.

\bigskip
{\bf 4. Conclusions}
\bigskip

 From all this we conclude that all hadronic total cross-section data are 
compatible with the stringy principles A) and B) above. 
The reason for the apparent discrepancy between the fits of references
\cite{DL, DG} and these principles is that for simplicity,
they i) ignored Regge-Regge cuts and 
ii) fit the Pomeron to a single power law.
The remarkable power of the principles A) and B) is that, even after
abandoning these simplifications, the necessary number of parameters 
did not increase. Actually, after all constraints were met, 
we found this number to have {\em decreased}.

In a forthcoming paper we shall further explore these principles in the light
of recent developments in open string theory. 

\newpage
{\bf Acknowledgment}
\bigskip

This paper was supported in part by NSF grant PHY-9123780-A3.

\end{document}